\newif\ifdouble
\newif\ifsingle
\newif\ifchange
\doubletrue

\ifdouble
  \documentclass[sigconf, usenames, dvipsnames]{acmart}
\else
  \documentclass[manuscript]{acmart}
  \singletrue
\fi

\usepackage{dblfloatfix} 
\usepackage{booktabs} 
\usepackage{arydshln}
\usepackage{subfig}
\newcommand{\subsub}[1]{\noindent\textit{\textbf{#1:}}}

\AtBeginDocument{%
  \providecommand\BibTeX{{%
    \normalfont B\kern-0.5em{\scshape i\kern-0.25em b}\kern-0.8em\TeX}}}

\copyrightyear{2023}
\acmYear{2023}
\setcopyright{acmlicensed}\acmConference[CHI '23]{Proceedings of the 2023 CHI Conference on Human Factors in Computing Systems}{April 23--28, 2023}{Hamburg, Germany}
\acmBooktitle{Proceedings of the 2023 CHI Conference on Human Factors in Computing Systems (CHI '23), April 23--28, 2023, Hamburg, Germany}
\acmPrice{15.00}
\acmDOI{10.1145/3544548.3581381}
\acmISBN{978-1-4503-9421-5/23/04}

\newcommand{\system}{ChameleonControl}

\begin{document}
\pagenumbering{arabic}
\pagestyle{plain}
\title{\system{}: Teleoperating Real Human Surrogates through Mixed Reality Gestural Guidance for Remote Hands-on Classrooms}

\author{Mehrad Faridan}
\affiliation{%
  \institution{University of Calgary}  
  \city{Calgary}
  \country{Canada}}
\affiliation{%
  \institution{MaKami College}
  \city{Calgary}
  \country{Canada}}  
\email{mehrad.faridan1@ucalgary.ca}

\author{Bheesha Kumari}
\affiliation{%
  \institution{University of Calgary}
  \city{Calgary}
  \country{Canada}}
\email{bheesha.kumari@ucalgary.ca}

\author{Ryo Suzuki}
\affiliation{%
  \institution{University of Calgary}
  \city{Calgary}
  \country{Canada}}
\email{ryo.suzuki@ucalgary.ca}

\renewcommand{\shortauthors}{Faridan, Kumari, and Suzuki.}
\begin{abstract}
We present ChameleonControl, a \textit{\textbf{real-human teleoperation}} system for scalable remote instruction in hands-on classrooms. In contrast to existing video or AR/VR-based remote hands-on education, ChameleonControl uses a \textit{real human} as a surrogate of a remote instructor. Building on existing human-based telepresence approaches, we contribute a novel method to \textit{teleoperate} a human surrogate through synchronized mixed reality hand gestural navigation and verbal communication. By overlaying the remote instructor's virtual hands in the local user's MR view, the remote instructor can guide and control the local user as if they were physically present. This allows the local user/surrogate to synchronize their hand movements and gestures with the remote instructor, effectively teleoperating a real human. We deploy and evaluate our system in classrooms of physiotherapy training, as well as other application domains such as mechanical assembly, sign language and cooking lessons. The study results confirm that our approach can increase engagement and the sense of co-presence, showing potential for the future of remote hands-on classrooms.

\end{abstract}

\ccsdesc[500]{Human-centered computing~Mixed / augmented reality}

\keywords{Mixed Reality, Visual Cue, Remote Collaboration, Telepresence, Remote Guidance, Human Surrogates, Hands-on Training}

\begin{teaserfigure}
\centering
\includegraphics[width=\linewidth]{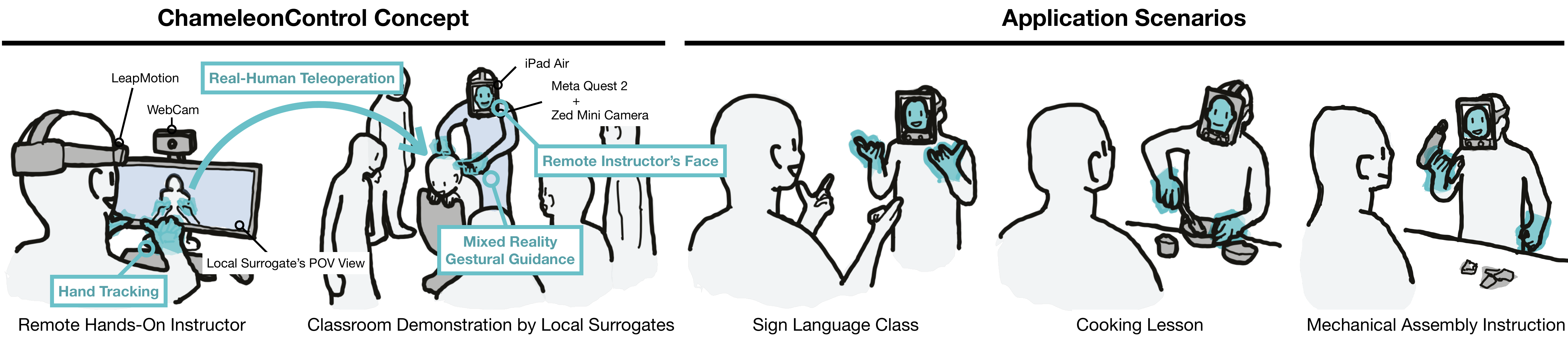}
\caption{\system{}, a real-human tele-operation concept through mixed reality gestural guidance. A remote instructor demonstrates a hands-on lesson to a classroom of students via a synchronized real-human surrogate (left) with additional possible hands-on application scenarios (right).}
\label{fig:teaser}
\Description{This sketched image consists of a left and right section. Displayed on the left is the ChameleonControl concept, where arrows and labels indicate what equipment is being used in the system and why. The webcam used to capture the instructor's face, head-mounted leap motion controller and the instructors monitor displaying a surrogate's first person point-of-view are displayed next to a human surrogate who is at a different location from the remote instructor and is shown from a third-person perspective while wearing the ChameleonControl custom HMD, with labels indicating the Quest 2, Zed Mini Camera and iPad Air and performing the remote instructor's classroom physiotherapy demonstration for a student audience. The second part of the image, on the right, shows the 3 non-physiotherapy application scenarios, namely a sign language lesson (where a surrogate is demonstrating some sign language word), a cooking lesson (where a surrogate is demonstrating whisking) and mechanical assembly instruction (where a surrogate is holding a drill)}
\end{teaserfigure}

\maketitle

\section{Introduction}
Despite the widespread adoption of remote education, physical hands-on training and instruction still remain difficult to effectively teach or learn in a remote setting. 
In hands-on training, such as physical therapy training, mechanical task instruction, cooking lessons, and sports coaching, the physical demonstrations are vital to effective training, but such physical demonstrations cannot be captured well with a simple 2D video call like Zoom~\cite{hawkins2016limitations}.
To address this, recent research has explored mixed reality (MR) interfaces for live instruction (e.g., \textit{Loki}~\cite{thoravi2019loki}) or recorded video tutorials (e.g., \textit{ProcessAR}~\cite{chidambaram2021processar}, \textit{AdapTutAR}~\cite{huang2021adaptutar}) for hands-on training and education.
These mixed reality interfaces show great promise to improve the sense of co-presence and the ability to demonstrate such lessons in a more spatial and immersive manner~\cite{cao2020exploratory}.

However, this approach \textit{does not scale well for classrooms}, where dozens of students learn by observing an instructor's hands-on demonstration.
In such a case, every student needs to wear a MR headset, which incurs a large financial cost as well as a cumbersome preparation and calibration process that significantly hinders practical use in actual classrooms.
Moreover, even if the headset becomes cheap, small, and easily accessible, the \textit{virtual} demonstration is still far from the actual \textit{physical} demonstration, especially when the embodied interactions matter~\cite{ferdous2019s}. 
For example, we learn from our formative study that the lack of physical demonstrations and human touch results in a poor learning experience in certain domains like physiotherapy training.
Because of that, many kinds of classroom-scale hands-on training still need to rely on physical and co-located instruction~\cite{hoang2017augmented}, as it is difficult to deliver an adequate educational experience in remote settings.

In this paper, we propose a novel approach to remote classroom instruction that uses a \textit{real human} to physically demonstrate a remote instructor's lesson for classroom-scale hands-on training. By equipping a human surrogate with an MR headset and a remote instructor with a head-worn hand-tracking device (Leap Motion), the remote instructor's hands can be overlayed within the surrogate's MR point-of-view, enabling the surrogate to synchronize their real hands with the remote instructor's virtual hands via MR gestural guidance and verbal instruction. This enables the remote instructor to \textit{physically} demonstrate hands-on lessons via the guided human surrogate at classroom scale. Additionally, we modify the surrogate's MR headset with a face-mounted 2D tablet display to display the remote instructor's face on top of the surrogate's face in order to enhance the sense of co-presence of in-person students with the remote instructor.

Our approach takes inspiration from the existing human-surrogate telepresence (e.g., \textit{ChameleonMask}~\cite{misawa2015chameleonmask}), but, we offer two novelties built on top of it: 
\subsub{1) Teleoperate, not telepresence} First, we leverage mixed reality virtual hand overlays to \textit{teleoperate} a human surrogate. In contrast to the existing voice-based~\cite{misawa2015chameleonmask, yamada2022and} or 2D finger pointing-based navigation~\cite{misawa2015wearing, misawa2016touching}, 3D hand gestural guidance enables more precise synchronized body motion, as if the remote instructor is teleoperating the local surrogate.
\subsub{2) Re-purpose the local user/surrogate as a classroom instructor} Second, we explore and evaluate the use of the local user as a classroom instructor for hands-on training. With this, our approach can be used not only for one-on-one training (navigating a local surrogate through MR) but more importantly for scalable hands-on/physical demonstrations in classrooms of co-located students (i.e. letting other students observe the hands-on and physical instructions or lesson given by a remote instructor by watching the co-located and remotely synchronized/teleoperated surrogate).
To demonstrate this concept, we develop \system{} system, which leverages a LeapMotion for the remote instructor's hand tracking and a custom MR headset based on Meta Quest 2, Zed Mini camera, and iPad for the local surrogate user.

To evaluate our system, we conducted two user studies evaluating the student audience, co-located surrogate, and remote instructor's experiences with \system{} as compared to other remote teaching methods (e.g., Zoom) and a final third user study to evaluate \system{} in other application domains, such as mechanical assembly, sign language, and cooking lessons.
From our study, we found that our approach is effective for teaching and learning hands-on training in various application domains.
Based on the findings, we discuss the ethics of using human surrogates, the uncanny valley, and how we can generalize this concept for the future of real-human teleoperation.

\begin{figure*}
\centering
\includegraphics[width=\textwidth]{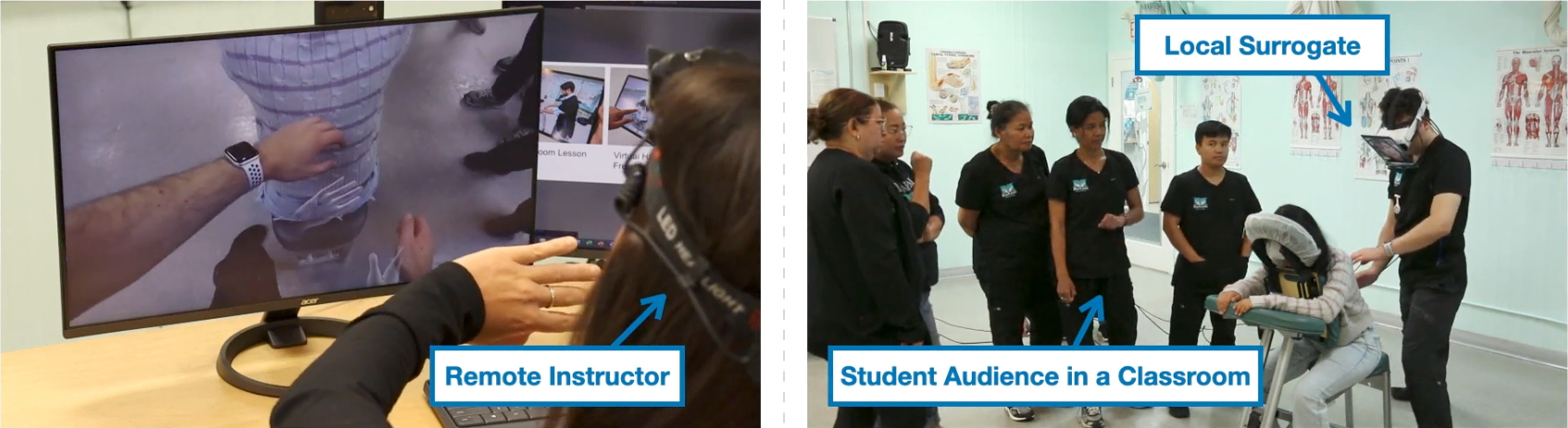}
\caption{A perspective overview of \system{}. The remote instructor views the surrogate's Point-of-View and guides them using hand gestures (left) while the local surrogate synchronizes for the student audience (right).}
\label{fig:overview}
\Description{This image consists of a left and right section. On the left is a real image of a remote physiotherapy instructor's POV watching the human-surrogate's streamed POV on a desktop monitor. On the right is an image showing 5 real massage therapy students gathered watching the human surrogate perform the lesson, in this case a chair massage lesson on a non-student volunteer participant, while equipping the ChameleonControl custom HMD.}
\end{figure*}

Finally, our contributions are the following: 
\begin{enumerate}
\item The concept of real-human teleoperation by combining human-surrogate telepresence and mixed reality gestural guidance. 

\item \system{}, a system that enables remote and scalable teaching/instruction for hands-on training in classrooms, with design decisions and our motivation informed by our formative study.

\item Three user studies to measure the effectiveness of our approach in various hands-on training scenarios and from each user's perspective (instructor, surrogate, and student audience).

\end{enumerate}

\section{Related Work}

\subsection{Mixed Reality for Hands-on Training}
Mixed reality provides an effective means for hands-on training for physical tasks~\cite{wang2021ar, baroroh2021systematic}, such as mechanical inspection (e.g., \textit{ARMAR}~\cite{henderson2007augmented}), machine task (e.g., \textit{AdapTutAR}~\cite{huang2021adaptutar}, \textit{MobileTutAR}~\cite{cao2022mobiletutar},  \textit{ProcessAR}~\cite{chidambaram2021processar}), assembly task (e.g., \textit{SHARIDEAS}~\cite{wang2021sharideas}, \textit{Teach Me How}~\cite{funk2018teach}), exercise instruction (e.g., \textit{My Tai Chi Coaches}~\cite{han2017my}), and music lessons (e.g., \textit{Loki}~\cite{thoravi2019loki}). 
In fact, Cao et al.~\cite{cao2020exploratory} show that the human avatar in these tutorials significantly improves the spatial and body-coordinated interaction than the no-avatar condition. 

However, there are still some challenges and limitations in MR hands-on training.
First, HMD-based approaches are still difficult to seamlessly integrate into the physical world. 
For example, Funk et al.~\cite{funk2016interactive} reveals that projection mapping enables faster task completion time and less cognitive load than HMDs in block assembly tasks, which indicates that the low level of integration of HMDs significantly diminishes the effectiveness of instruction.
Second, it is difficult to apply in classroom settings due to the scalability problem.
In fact, most of the existing works focus on \textit{one-on-one} instruction~\cite{thoravi2019loki,chidambaram2021processar,cao2020exploratory}, but hands-on demonstrations are also important for \textit{one-to-many} scenarios like physiotherapy classrooms~\cite{ferdous2019s}.
Existing works for classroom hands-on education either focus on projection mapping in co-located settings (e.g., \textit{Augmented Studio}~\cite{hoang2017augmented}, \textit{HoloBoard}~\cite{gong2021holoboard}, \textit{HOBIT}~\cite{furio2017hobit}) or fully virtual environments in VR/AR for remote instruction (e.g., \textit{XRStudio}~\cite{nebeling2021xrstudio}, \textit{AR/VR for Liver Anatomy Education}~\cite{schott2021vr}, \textit{RealityTalk}~\cite{liao2022realitytalk}).
We address this by enabling remote hands-on instruction for classroom settings by exploring real-human teleoperation.

\subsection{Telepresence}

\subsubsection{Video-based Telepresence}
Video-based telepresence has been studied for decades~\cite{ishii1992clearboard, gauglitz2014touch}.
Due to the COVID-19 pandemic, video-based telepresence (e.g., Zoom, Microsoft Teams) has become almost ubiquitous.
However, these screen-based telepresence systems are still limited for hands-on instruction due to the lack of spatial and embodied interactions~\cite{hawkins2016limitations}.

\subsubsection{Virtual Telepresence}
By leveraging the recent advances of AR/VR devices, many works have explored virtual telepresence through 3D teleportation (e.g., \textit{Holoportation}~\cite{orts2016holoportation}, \textit{Virtual Makerspaces}~\cite{radu2021virtual}), avatars (e.g., \textit{CollaboVR}~\cite{he2020collabovr}, \textit{Mini-Me}~\cite{piumsomboon2018mini}, \textit{Loki}~\cite{thoravi2019loki}), and projected images (e.g., \textit{Room2Room}~\cite{pejsa2016room2room}).
These approaches allow more immersive remote collaboration, as the virtual hand and body representation can provide a significantly stronger sense of co-presence than without it~\cite{bai2020user,cao2020exploratory}. 
These virtual telepresence systems can be also used for remote instruction of physical tasks (e.g., \textit{Loki}~\cite{thoravi2019loki}, \textit{CollabAR}~\cite{villanueva2022colabar}, \textit{ARTEMIS}~\cite{gasques2021artemis}).
However, virtual telepresence lacks physical embodiment, which limits the ability to collaborate through physical objects or bodies~\cite{leithinger2014physical, rae2014bodies}.

\subsubsection{Robotic Telepresence}
HCI researchers have also explored \textit{physical telepresence}~\cite{leithinger2014physical}, as opposed to virtual telepresence by leveraging robotic telepresence (e.g., \textit{TELESAR V}~\cite{fernando2012design}, \textit{MeBot}~\cite{adalgeirsson2010mebot}, \textit{Telenoid}~\cite{ogawa2011exploring}, \textit{You as a Puppet}~\cite{sakashita2017you}, \textit{GestureMan}~\cite{kuzuoka2000gestureman}, \textit{Geminoid}~\cite{sakamoto2007android}) or actuated tangible interfaces (e.g., \textit{TRANSFORM}~\cite{leithinger2014physical}).
Prior work has shown that the physical actions of these robots, such as movement~\cite{nakanishi2011zoom}, mobility~\cite{rae2014bodies, lee2011now} and gestures~\cite{adalgeirsson2010mebot} can greatly improve engagement and social interactions.
Such robotic telepresence is also used in the context of education (e.g., \textit{RobotAR}~\cite{villanueva2021robotar}, \textit{ASTEROIDS}~\cite{li2022asteroids}). However, these robots mimic human appearance and motion very poorly.
While some works like \textit{VROOM}~\cite{jones2020vroom, jones2021belonging} or \textit{LiveMask}~\cite{misawa2012livemask} can augment the robot's appearance through \textit{``virtual skins''} of humans, it can only mimic visual aspects but not physical embodied interaction.
The works like \textit{Parallel Ping-Pong} ~\cite{takada2022parallel} demonstrate a high degree of freedom to mimic a remote user's motion, but it is still very difficult to simulate a complex embodied interaction, which is a requirement of most hands-on training such as mechanical tasks or physical therapy training.

\subsubsection{Human-Based Telepresence}
To address this, human-based telepresence has been explored as an alternative approach.
Human-based telepresence was originally developed as an idea of using a remote user as a mobile camera to experience the same interaction remotely (e.g., \textit{Tele-Actor}~\cite{goldberg2002collaborative}, \textit{TEROOS}~\cite{kashiwabara2012teroos}, \textit{JackIn Head}~\cite{kasahara2015jackin}, \textit{Shopping Together}~\cite{cai2018shopping}, \textit{Go Together}~\cite{cai2019go}). 
Going beyond that, researchers started exploring the idea of real human telepresence by showing the remote user's faces on top of the local actor~\cite{misawa2015wearing}.
Most closely related to our work, \textit{ChameleonMask}~\cite{misawa2015chameleonmask} uses a human surrogate for human-based telepresence, which can be used for various applications such as shopping~\cite{misawa2015wearing} and 
theater performance~\cite{tholander2021design}.
In terms of controlling the local user, however, these works only leverage voice-based~\cite{misawa2015chameleonmask} or finger-pointer-based commands~\cite{misawa2016touching, misawa2015wearing}, and no prior work, to the best of our knowledge, has investigated 
the immersive gestural guidance, which enables scalable remote instruction for hands-on and physical education.

\subsection{Visual Cues for Remote Guidance}
Many researchers have investigated the use of visual communication cues to guide remote users for spatial and physical tasks~\cite{fussell2000coordination, fussell2003effects}.
These communication cues can take various forms, such as sketches~\cite{gauglitz2014world, gauglitz2014touch, jalaliniya2014wearable, huang2019sharing, teo2019investigating, gao2020user}, pointers~\cite{lee2020enhancing, kim2020use}, gaze~\cite{gupta2016you, higuch2016can}, hands~\cite{kuzuoka1994gesturecam, huang2018augmented}, and combinations of the above~\cite{kim2019evaluating, kim2020combination, higuch2016can, chen2013semarbeta}.
For example, sketch-based visual cues are often used for a spatial annotation for real-time remote guidance (e.g., \textit{TransceiVR}~\cite{thoravi2020transceivr}, \textit{SEMarbeta}~\cite{chen2013semarbeta}).
Alternatively, pointer or gaze cues are used to show where the remote user should look at in task instruction (e.g., \textit{AR Tips}~\cite{lee2020enhancing}, \textit{Do You See What I See}~\cite{gupta2016you}, \textit{Can Eye Help You}~\cite{higuch2016can}).
Different modalities such as vibro-tactile feedback (e.g., \textit{HapticPointer}~\cite{matsuda2020hapticpointer}) or electrical muscle simulation (e.g., \textit{BioSync}~\cite{nishida2017biosync}, \textit{Paralogue}~\cite{hanagata2018paralogue}) have also been explored as an implicit guide. 
One of the most common visual communication cues is hand gestures (e.g., \textit{Augmented 3D Hands}~\cite{huang2018augmented},  \textit{GestureCam}~\cite{kuzuoka1994gesturecam}, \textit{RemoteFusion}~\cite{adcock2013remotefusion}, \textit{Turn It This Way}~\cite{kirk2007turn}, \textit{MirrorTablet}~\cite{le2017mirrortablet}, \textit{Imitative Collaboration}~\cite{zhang2022imitative}).
Similar to our work, prior works use virtual hands for remote guidance shown in MR headset~\cite{kim2019evaluating, teo2018hand} to navigate novice users for various physical tasks, such as block assembly~\cite{zhang2022real}, origami~\cite{kim2019evaluating, kim2020combination}, and mechanical tasks~\cite{oyama2021augmented, oyama2021integrating}.
Wang et al.~\cite{wang2022evaluating} further investigate the use of gestural cues for one-to-many remote collaboration.
While this MR-based gestural guidance itself is not new, our novelty lies in the first combination of these gestural visual cues with the human-surrogate telepresence to achieve the real-human teleoperation concept for scalable remote instruction for hands-on classrooms, which has never been explored nor evaluated in the literature.

\section{\system{}}

\subsection{Overview}
This section introduces \system{}, a system that enables scalable remote instruction for physical, hands-on, and spatial lessons in classrooms by using teleoperated real-human surrogates.

\system{} streams a first-person point-of-view at low latency (30-40ms) video passthrough from the local surrogate's stereoscopic camera (ZED Mini) to both the local surrogate's equipped custom HMD and a remote user's desktop. Both the local surrogate's custom HMD and stereoscopic camera are connected to a PC via USB-C/USB-A cables. The video feed is streamed stereoscopically to the local surrogate's custom HMD enabling them to naturally and immersively interact with their physical environment. The video passthrough is streamed monoscopically to a remote user's desktop, enabling the remote user to view the scene from the local surrogate's 2D first-person point-of-view.

On the other hand, \system{} uses a local surrogate not only as a camera streamer but also as a surrogate of the remote user, by displaying the facial expressions of the remote user as a \textit{``mask''} of a local surrogate.
All in all, the local surrogate's custom HMD consists of: 1) Meta Quest 2 VR headset, 2) iPad display mounted to the front of the headset, 3) ZED Mini stereoscopic camera mounted to the front of the iPad. The remote user equips a simple head strap with a LeapMotion attached, connected via wire to a PC.
In addition, \system{} also displays a remote user's hands in a local surrogate's mixed reality first-person camera point of view (POV), so that the remote user can guide how the local surrogate should act and behave via navigated hand gestures.

By synchronizing both facial expressions and bodily motion, \system{} creates an illusion, as if the remote user is \textit{teleoperating} a local surrogate, whose body motion is visible to a co-located audience. 
Figure~\ref{fig:teaser} illustrates the overall system design.

\subsection{Formative Study and Design Rationale}
The design of our system is informed by an informal formative interview with six experts who provide classroom hands-on instruction in various domains---three physical therapy instructors, one professional yoga instructor, one sports coach, and one high-school science teacher who conducts a science experiment.
During one hour interview, we first asked open-ended questions about their needs, opportunities, and challenges for remote teaching. Then, we also asked how they perceive the current remote teaching approaches like zoom video and virtual telepresence~\cite{orts2016holoportation, thoravi2019loki} by showing an image and video for context and inspiration. 

\subsubsection{Challenges of Remote Hands-on Training}
When we asked about the possibility of remote instruction, most of them express significant challenges.
For example, experts mention that remotely teaching students through Zoom is not as effective as co-located instruction, as students cannot see from different angles, and the instructor also cannot observe students, which diminishes the coaching experience.
Although some experts see promise in the MR system, other experts doubt its effectiveness due to fundamental differences between co-located and virtual experiences.
For example, in certain areas like physical therapy, they emphasize the importance of physical demonstration, as the lack of physical demonstration and human touch leads to a poor learning experience.

\subsubsection{Scalability for Classroom Uses}
While experts saw potential benefits of adopting virtual 3D remote instruction like \textit{Holoportation}~\cite{orts2016holoportation}, another challenge expressed by experts is the practicality and scalability of these AR/VR-based approaches.
For example, they express a reluctance to adopt HMDs at scale in their classrooms due to accessibility issues.
Affordability is another issue, as it requires a prohibitive cost.
In addition, they express concern that their demonstrations would not be clearly and accurately visualized, especially when touching or interacting with physical bodies.
Overall, they see benefits for the MR system for one-on-one instruction, but it is not clear how it can be used in classrooms.

\subsubsection{Strong Needs and Opportunities for Remote Instruction}
On the other hand, they also express a strong need for and interest in remote education.
For example, some of the experts live far from their classroom studios and this constant long-distance travel often causes significant logistical and scheduling problems.
Additionally, they also express great interest in further benefits of remote teaching, in particular the ability to scale their lessons to reach as many students as possible as easily as possible.

\subsubsection{Design Goals and Rationale}
In summary, through the formative interviews, we learned
\begin{enumerate}
\item The observation of the instructor's demonstration is vital for classroom hands-on training
\item There is a strong need for remote instruction but current methods are not suitable
\item Mixed reality instruction is promising for one-on-one training, especially with the first-person POV demonstrations and coaching.  
\item The method should be scalable, accessible, and affordable for use in classroom settings for each student.
\end{enumerate}

Based on these requirements, we gradually developed a basis for \system{}'s approach, which re-purposes a real human surrogate as a physical demonstrator.
With this approach, we solve two problems at once: 1) providing one-on-one demonstration through mixed reality feedback for a local student, and 2) using the local student as a physical demonstrator for other students in the classroom.

\subsection{System and Implementation}
In \system{}'s framework, there are three types of users: 1) a remote instructor, 2) a local surrogate, and 3) a student audience in a classroom.

\subsubsection*{1) Remote Instructors}
First, a remote instructor is an expert who teaches and demonstrates hands-on training to the student audience, such as physical therapy training, yoga and exercise instruction, mechanical assembly, science experiments, sports training, cooking lessons, and other hands-on physical tasks. 
In our system, the remote instructor can watch the local surrogate's first-person POV through her head-worn camera (Zed Mini) which streams to a 2D display (Acer 24-inch display) in front of the instructor.
On the other hand, the instructor's face captured by a webcam (Logitech C930) is shown in the local surrogate's face through a video conferencing application (Google Meet).
On top of that, the remote instructor's hand movement is captured through a hand motion tracking device (Leap Motion), which can cover a wide 80cm by 60cm tracking area 120 degrees vertically and 140 degrees horizontally.
Both the instructor and the local surrogate can see the overlaid virtual hands in their POV, which allows synchronized body motion between the remote and local surrogates. Additionally, the instructor can communicate verbally with the surrogate via the iPad-face video conferencing call or an audio call on a separate nearby device (laptop, phone, etc) in cases when the iPad-face wasn't used.
The tracked hand skeleton information is provided by the Leap Motion to Unity application, which is run on a Windows machine (OS: Windows 11, CPU: Intel Core i7-11700K, GPU: Nvidia RTX 3050, RAM: 32GB Crucial Ballistix) with 60 FPS.
The remote instructor can see the local surrogate's camera view, in which their own virtual hands (Ultraleap's Ghost Hands model/texture) are also overlaid on top of the camera view.

\subsubsection*{2) Local Surrogates}
Second, a local surrogate acts as a physical embodiment of the remote instructor. 
The local surrogate wears a custom MR headset, which consists of three components: 1) \textbf{Headset:} Meta Quest 2 for a mixed reality headset, 2) \textbf{Display:} an iPad Air for a display screen to show the remote instructor's face, 3) \textbf{Camera:} a Zed Mini stereoscopic camera to capture the outside view, effectively converting the Meta Quest 2 from a VR to an MR display.
We mount all of these components with a 3D-printed custom mount and Velcro tape.
First, the attached stereoscopic camera captures the local person's view which is streamed to VR headset, so that the local surrogate can see as if they're looking through a pass-through display. 
The same camera is also streamed to the remote instructor, so that they can see the same POV at the same time.
The remote instructor's 3D gestural navigation is overlaid on top of the stereoscopic camera feed as a pair of virtual hands (for both local and remote users' scenes).
While Meta Quest itself has a grayscale passthrough view, we cannot use this because the built-in cameras are occluded with the iPad screen. 
For the same reason, we avoid all built-in passthrough cameras (e.g., Vive Pro) and instead use an external camera like the Zed Mini, which can be attached on top of the iPad screen.
Zed Mini is connected to a local Windows machine (OS: Windows 11, CPU: Intel Core i7-11700K, GPU: Nvidia RTX 3050, RAM: 32GB Crucial Ballistix) via a cable.
The machine renders the captured 3D scene in the Unity application, then streams it to the local surrogate's Quest through Oculus Link and SteamVR. As mentioned before, the surrogate can communicate with the remote instructor verbally via the iPad-face video conferencing call or via an audio call on a separate nearby device (laptop, phone, etc) in cases when the iPad-face wasn't used.
The total weight of the custom headset (Meta Quest 2, Additional Battery Pack, iPad Air 10-inch, and Zed Mini) is 1.30 kg.

Since the role of the local surrogate is to act and behave exactly like the remote instructor does, the local surrogate is supposed to move her hands by matching with the overlaid virtual hands, so that their body motions can be synchronized.
Moreover, the attached tablet display shows the remote instructor's face to create an expression and illusion that the remote instructor is co-located.
The local surrogate can be anybody capable of synchronizing with the remote user, the requirements for which depend on the use case. For example, some situations may not require surrogates to move around (i.e., surrogates can be seated), while others may only require synchronization of one hand. In general, we believe that anyone able-bodied and capable of synchronizing properly with the remote user can play the role of a surrogate. We envision that in classroom settings, a willing student volunteer or teaching assistant will likely play the role of a surrogate.

\subsubsection*{3) Student Audience in a Classroom}
Finally, there is a student audience that observes the local surrogate's demonstration.
In contrast to the existing works (e.g., ChameleonMask~\cite{misawa2015chameleonmask, misawa2015wearing, misawa2016touching}), our focus is specifically on remote hands-on training for classrooms. 
In our system, the local surrogate herself provides the same demonstration for students, since the local surrogate's and remote instructor's body motions should be synchronized.
Unlike other mixed reality hands-on training systems~\cite{thoravi2019loki, huang2021adaptutar, chidambaram2021processar}, the student audience does not need any equipment such as MR/VR head-mounted displays (e.g., Hololens) or mobile AR devices, as they can observe the hands-on physical demonstration in the same way as they do in co-located hands-on instruction. Finally, the student audience can communicate verbally with the remote instructor via the iPad-face video conferencing call or via an audio call on a separate nearby device (laptop, phone, etc) in cases when the iPad-face was not used.

\section{User Study}

\subsection{Overview of Three User Studies}
To evaluate \system{}, we conducted the following three different user studies: 

\subsub{1) In-Class Deployment for Physiotherapy Training} In this study, we evaluate our system by deploying in a hands-on training classroom of physio and massage therapy education, where a total of 20 students experience and compare our system with different lesson-delivery methods as a classroom audience.

\subsub{2) Expert Review from Physical Therapy Instructors and Students} 
Moreover, we conduct an in-depth user study with 10 instructors and 14 students (split into pairs with one instructor and one student each) in order to evaluate our system from the remote instructor and local student surrogate perspectives.

\subsub{3) Usability Test for Different Applications}
This study evaluates the effectiveness of our system for different applications beyond the previous studies' specific domain of physiotherapy training.
To this end, we prepare three sample hands-on instruction tasks in cooking lessons, mechanical assembly training, and sign language instruction as possible hands-on classroom situations and evaluate with 9 participants from the local surrogate and student audience perspectives.

\subsection{In-Class Deployment for Physiotherapy Classrooms}

\subsubsection{Research Questions}
For this study, our goal was to evaluate the \textbf{student audience's experiences} in classrooms as compared to the other approaches. 
To achieve this, we designed our study with 4 distinct lesson-delivery methods, detailed as follows:

\begin{itemize}
\item \textbf{Zoom:} Remote teaching based on Zoom. 
\item \textbf{MobileAR:} Virtual hand avatars shown on iPad screen
\item \textbf{No Face:} \system{} without the iPad face
\item \textbf{Ours:} \system{} 
\end{itemize}
We prepared these conditions because we are interested in evaluating the following research questions: 

\begin{itemize}
\item \subsub{RQ1. Physical vs Virtual Demonstration} Compared to Zoom or the Mobile AR hand avatar condition, does the in-person physical demonstration by the real-human surrogate improve the learning experiences?
\item \subsub{RQ2. Face vs No Face} Compared to the no face condition, does the face shown on top of the local surrogate (via their face-mounted iPad) improve the audience's sense of co-presence with the remote instructor?
\end{itemize}

\subsubsection{Method and Participants}
To evaluate our system in an actual hands-on classroom, we deploy our system to a professional massage therapy school, the largest massage therapy school in our local community~\footnote{\url{https://makamicollege.com/}}. 
We recruited a total of 20 participants (9 male, 11 female, ages 18-49) to evaluate the classroom audience perspective.
All of the participants are actual students in the school who are learning to become licensed therapists.
The study was deployed in one of the classrooms in their school and conducted over 4 days.
Participants were first introduced to the overview of our study and then experienced the four conditions one by one.
We had a total of four different sessions (each session had roughly 5 participants), with the order of the conditions randomized for each session.

For the remote instructor and local surrogate, we recruited professional physiotherapy instructors from the same school for each session.
For the \textbf{Zoom condition}, the instructor's demonstration is streamed from a third-person view, with the camera angle controlled by an author.
For the \textbf{mobile AR virtual hand condition}, we show the instructor's real-time hand movement tracked with a Leap Motion Controller, then overlay it on top of the scanned 3D patient in a Unity environment. 
Each student can then watch the virtual hand movement in free-viewpoint with the iOS "Unity Virtual Camera" app on an iPad and can control the camera position displayed on the iPad by tilting it as well as using two control inputs, one as a joystick for forward/backward movement as well as left/right movement. The other input control allows users to move the camera up and down. This way, users have a full 6-degrees of freedom to move around the unity scene.
We needed to choose this method as using an iPad is the only scalable way for each student to experience real-time virtual telepresence.
For the \textbf{no face condition}, we use the same \system{} system, but without the face-mounted iPad (just the Meta Quest 2 headset and attached Zed Mini camera), so that the audience doesn't see the instructor's face overlayed on the surrogate's face. 
For each condition, the remote instructor provides a real chair massage therapy lesson as they do in a regular classroom setting, and after 5-10 minutes, we switch to the next condition.
Once the participants experienced all conditions, then they were asked to fill out an online questionnaire form.
In total, the study took approximately 60 minutes and each participant was compensated 10 CAD.

\begin{figure*}
\centering
\includegraphics[width=\textwidth]{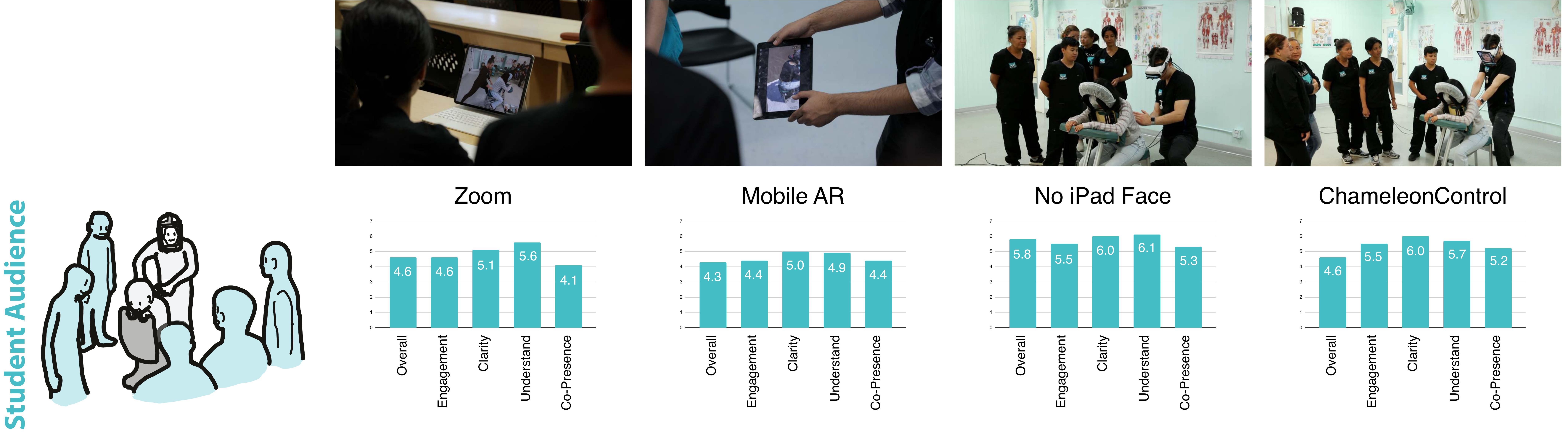}
\caption{The four conditions of the first user study deployed in-class for physiotherapy lessons and evaluated by the student audience.}
\label{fig:study-result-1}
\Description{This image displays a data visualization showing the results of the first user study which was conducted to assess the student audience's perspective on 4 different lesson-delivery methods: Zoom, Mobile AR, ChameleonControl without the iPad face and ChameleonControl.}
\end{figure*}

\subsubsection{Results}
Figure~\ref{fig:study-result-1} summarizes the 7-point Likert-scale questions for each condition. 
For Figure~\ref{fig:study-result-1}, ~\ref{fig:study-result-2} and ~\ref{fig:non-massage-study}, the actual questionnaire was ``Please rate from 1 to 7 (with 7 being the best) for the followings: 
1) \textbf{Overall}: Overall, how much you liked each lesson-delivery method, 
2) \textbf{Engagement}: How engaging you felt each lesson-delivery method was, 
3) \textbf{Clarify}: How clear and informative you felt each lesson-delivery method was, 
4) \textbf{Understand}: How easy-to-understand each lesson-delivery method was,
and
5) \textbf{Co-Presence}: How much you perceived the remote user's co-presence for each lesson-delivery method''.
These questions were inspired by and partially derived from the literature (e.g., co-presence:~\cite{kim2019evaluating, lee2017mixed, adalgeirsson2010mebot}, engagement:~\cite{adalgeirsson2010mebot}, easy to understand:~\cite{oyama2021integrating, lee2017mixed}, etc).
Below, we discuss the following aspects in more detail.

\ \\
\subsub{RQ1. Physical vs Virtual Demonstration}
When comparing our system with Zoom or Mobile AR instructions, our system improves the \textbf{level of engagement} over both Zoom and Mobile AR based on the average 7-point Likert-scale score (Ours: 5.5, Zoom: 4.6, Mobile AR: 4.4).
The participants also report that our system's \textbf{clarity and informativeness of instruction} is better than both Zoom and Mobile AR (Ours: 6.0, Zoom: 5.1, Mobile AR: 5.0). 
Participants responded positively to the physical and in-person experiences our system can provide.
For example, \textit{``P4: Real person with iPad face was better because I felt like the teacher was actually in person''}.
\textit{``P19: The VR (our system) felt more like I was in class learning than anything''}
\textit{``P17: Felt like a person was there (with our system)''}
The physical demonstration also allows the participants to see the instruction in much easier ways.
\textit{``P6: Seeing a person perform in person is easiest to view''}.

Participants also rated our system as a similar \textbf{overall experience} to Zoom and Mobile AR (Ours: 4.6, Zoom: 4.6, Mobile AR: 4.3), higher for the sense of co-presence (Ours: 5.2, Zoom: 4.1, Mobile AR: 4.4), and the \textbf{easiness of understanding} (Ours: 5.7, Zoom: 5.6, Mobile AR: 4.9). Participants generally preferred the physical demonstration over the video-based instruction or Mobile AR telepresence. 
\textit{``P3: I believe that the demonstration with the instructor having a face on iPad felt more realistic and that the teacher is more present. Loved that.''}
\textit{P15: It’s more interesting to see the instruction being performed physically}.
Interestingly, however, participants reported that the Zoom lesson is also easy to understand because of familiarity.
\textit{``P3: Zoom is easy to understand because I am used to it''}. 
For the same reason, Mobile AR virtual hands were not positively appreciated due to the lack of familiarity.
\textit{``P18: Each student has a different age group and different education level, and it takes a long time to learn and adapt''}.
In that sense, our system allows easy adaption as the students do not need to equip or learn anything, which indicates the easy deployment and adaption for classroom lectures. 
When asked whether they overall prefer our system, compared to zoom or mobile AR, participants responded (7: strongly agree, 1: strongly disagree) indicating that, overall, most students preferred our system to other methods.

\ \\
\subsub{RQ2. Face vs No Face}
In contrast to our expectations, overall, participants slightly prefer the no face condition for overall experiences (Ours: 4.6, No Face: 5.8), the sense of co-presence (Ours: 5.2, No Face: 5.3), easiness of understanding (Ours: 5.7, No Face: 6.1), the level of engagement (Ours: 5.5, No Face: 5.5), and clarity of the instruction (Ours: 6.0, No Face: 6.0).
However, some participants see the benefits of the visibility of the remote instructor's face. 
For example, 
\textit{``P10: Being able to see or even hear the instructor is beneficial for me personally''}.
But other participants do not feel the strong benefits of seeing the face on iPad, mostly because of the immaturity of the current technology or method of using iPad. 
\textit{``P11: iPad face is just weird''}.
\textit{``P6: My preference would be to utilize the headset but the displayed face on a real person is a bit creepy''}.
One observation we had was that the participants watch the instructor's demonstration from various different angles and the face shown on the iPad is not properly visible from side angles/views.
We also noticed that most of the participants focus on the instructor's hands rather than her face, thus, we learn that the participants largely don't care whether the remote instructor's face is presented or not, as long as they demonstrate in a co-located and physical manner. 
\textit{``P5: I don’t really need to see an instructor's face''}.
\textit{``P20: I don't think it's necessary to have an iPad face. I prefer a real person with VR''}.
As a follow-up question, we also brought a Microsoft Hololens 2 headset and informally ask whether the participants prefer the instructor's face on iPad over another person's face with a see-through Hololens headset, the majority of the participants prefer the see-through headset, even though they see the surrogate's face instead of the instructor's.
We did not implement the Hololens condition, thus we could not formally compare between these two but this indicates that for some situations like physiotherapy training, we could leverage a mixed reality \textit{see-through} display for real-human teleoperation instruction in the future. 

\subsection{Expert Review from Physiotherapy Instructors and Students}
\subsubsection{Research Questions}
For the second user study, our goal is to evaluate the user experiences from both the \textbf{remote instructor's and the local surrogate student's perspectives}, as opposed to the student audience perspective. 
We designed our study with the following three conditions 
\begin{itemize}
\item \textbf{First-Person Zoom:} Zoom remote instruction from the first-person point of view
\item \textbf{Third-Person Zoom:} Zoom remote instruction from the third-person point of view
\item \textbf{Mixed Reality Gesture:} Mixed reality gestural guidance overlaid on top of the student's first-person view with our system
\end{itemize}
We prepared for these conditions because we are interested in evaluating the following research questions: 

\begin{itemize}
\item \subsub{RQ3. Immersive vs Non-Immersive Guidance for Local Students} Compared to the zoom-based guidance (similar to the existing methods in ChameleonMask~\cite{misawa2015chameleonmask, misawa2015wearing, misawa2016touching}), can the mixed reality gestural guidance improve the learning experience for the local surrogate student? 
\item \subsub{RQ4. Immersive vs Non-Immersive Guidance for Remote Instructor} Can the mixed reality gestural guidance provide both a better teaching experience as well as an easier way to control and guide the surrogate for teleoperation? 
\end{itemize}

\subsubsection{Method and Participants}
We recruited 10 instructors (4 male, 6 female; age: 26-48) and 14 students (4 male, 10 female; age: 18-58) from the same physiotherapy training institutions.
The instructors are all professional and licensed therapists who have at least 4 years of teaching experience and regularly teach in-person massage and physical therapy classes with 10-40 students.
The students are massage therapy students at the same school. 
The study was conducted in the same classroom location as the first study. 
The participants were first introduced to an overview of our study, and then experience three different conditions, one with our system and another two with Zoom from both first-person and third-person perspectives, in a randomized order. 
For each condition, the instructor is told to teach as they would do in person, and the local surrogate student is supposed to behave and mimic the instructor's behavior as a teleoperated person by synchronizing their hand movements with the remote instructor's virtual avatar hands.
For the \textbf{first-person Zoom} condition, we mount the webcam to the instructor's forehead and stream it to an iPad, which is placed in front of the local student. 
For the \textbf{third-person Zoom} condition, an author controls the camera angle/perspective and streams it to an iPad, which is placed in front of the local student.
Each condition takes 5-10 minutes, then we switch to the next condition. After experiencing each condition, participants were asked to fill out an online questionnaire. In total, the study took approximately 40 minutes and each participant was compensated 10 CAD.

\begin{figure*}
\centering
\includegraphics[width=\textwidth]{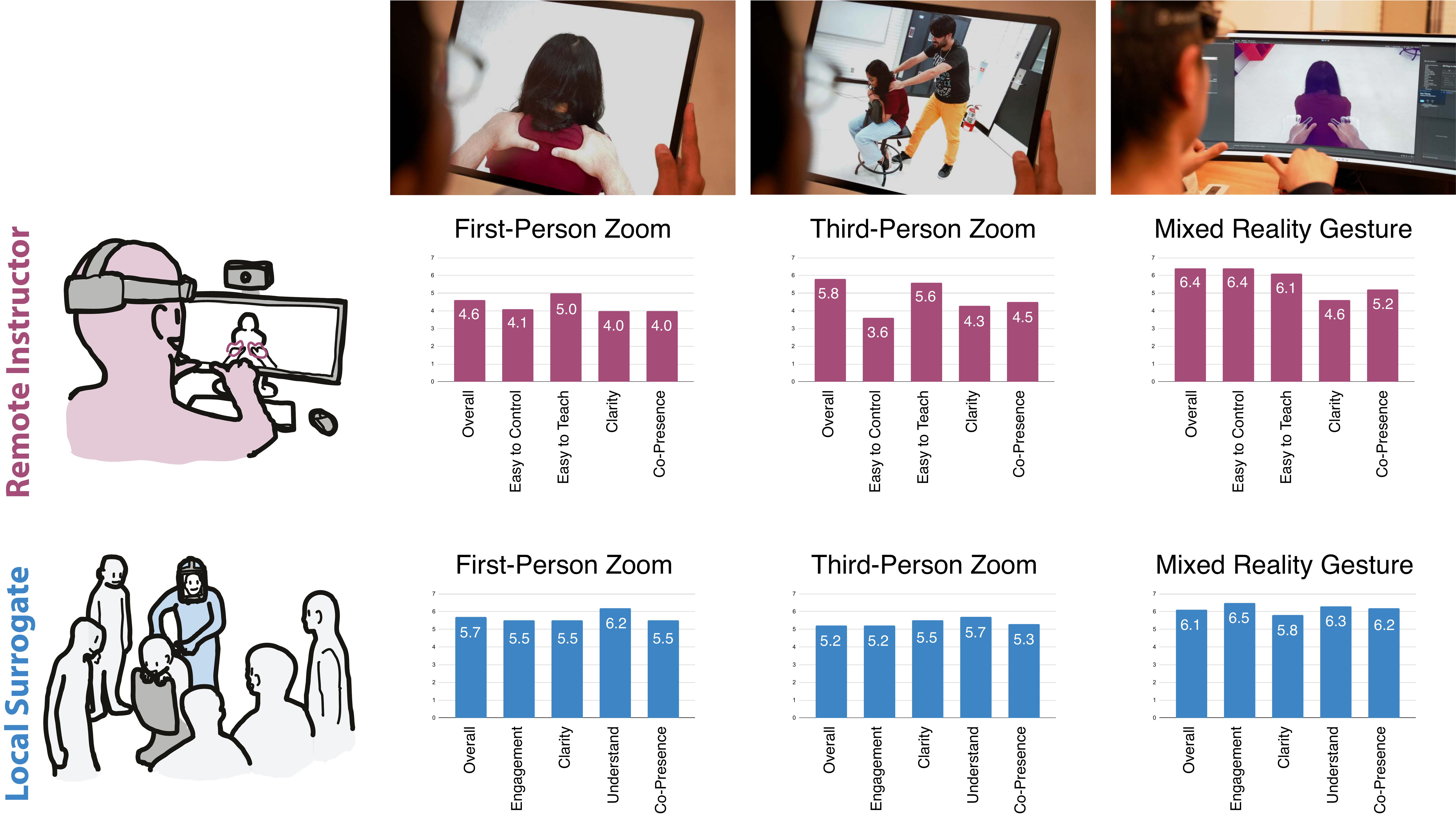}
\caption{The three conditions of the second user study deployed in-class for physiotherapy lessons and evaluated by the real-human surrogate and remote expert instructor.}
\label{fig:study-result-2}
\Description{This image displays a data visualization showing the results of the second user study which was conducted to assess the remote instructor and local surrogate perspectives on 3 different lesson-delivery methods: First-person Zoom, Third-person Zoom and ChameleonControl.}
\end{figure*}

\subsubsection{Results}
Figure~\ref{fig:study-result-2} summarizes the 7-point Likert-scale questions for each condition.
Below, we discuss the following aspects in more detail.

\ \\
\subsub{RQ3. Immersive vs Non-Immersive Guidance for Local Students}
When compared to our mixed reality gestural guidance with both Zoom-based methods, our system improves the \textbf{easiness of understanding} over both first-person POV and third-person POV video-based guidance (Ours: 6.3, 1st POV: 6.2, 3rd POV 5.7) for local surrogate students.
Many participants mentioned that the mixed reality approach is easy to follow (P2, P7, P9, P10) and easy to understand (P5, P7). 
\textit{``P5: With the immersive AR lesson the instructor's hands were easy to trace and steps were easy to replicate. 1st person wasn't as immersive and 3rd person was just bad''}.
\textit{``P6: The immersive AR was very practical for me, the 1st person zoom gave me more of a personal point of view perspective, but I still preferred the hands-on experience with the instructor as in the AR Lesson''}.
\textit{``P8: I can understand better how to position my hands and body''}. 
\textit{``P9: I could see how to position my hands After seeing my instructor's hands''}

The local surrogate students also report that the \textbf{level of engagement} was much better compared to the 1st-person POV and 3rd POV video-based guidance (Ours: 6.5, 1st POV: 5.5, 3rd POV: 5.2).
For example, participants mentioned the following
\textit{``P11: I really felt that the Vr experience was the most engaging''}.
\textit{``P5: immersive was really engaging because of the hand tracing 1st and 3rd person was just demonstrating''}.
\textit{``P2: AR- very engaging almost like a video game. It was a different feel and it kept my attention''}.
Participants also respond positively for other metrics, such as overall experiences (Ours: 6.1, 1st POV: 5.7, 3rd POV: 5.2), the sense of co-presence (Ours: 6.2, 1st POV: 5.5, 3rd POV: 5.3), and clarity and informative (Ours: 5.8, 1st POV: 5.5, 3rd POV: 5.5).  
In general, the third-person POV was the lowest rated for most of the metrics.
\textit{``P6: The 3rd person Zoom POV was my least favorite, just because with the other ones I could see more of the client I’m working on more closely''}.

On the other hand, some participants mentioned that the first-person Zoom had better visual clarity and relayed other contextual information better, such as how much pressure to be applied in massage training. 
\textit{``P2: For AR, the only thing was not knowing how much pressure to apply. On the other hand, the 1st video is much better to tell how much pressure to apply.''}
\textit{``P4: In AR even though you can see the instructor's hands it doesn't feel as if you are really with someone as the hands are very hard to make out sometimes where as in the videos you get to see the instructor fully''}.
The participants also mentioned that voice-based communication helped AR instruction a lot. 
\textit{``P4: The videos gave more visual clarity but the voice helped a lot in the AR''}.
Also, the participants mention that they need to have time to adapt. 
\textit{``P1: AR helped a lot. Just a little weird to get used to''}.

\begin{figure*}
\centering
\includegraphics[width=\textwidth]{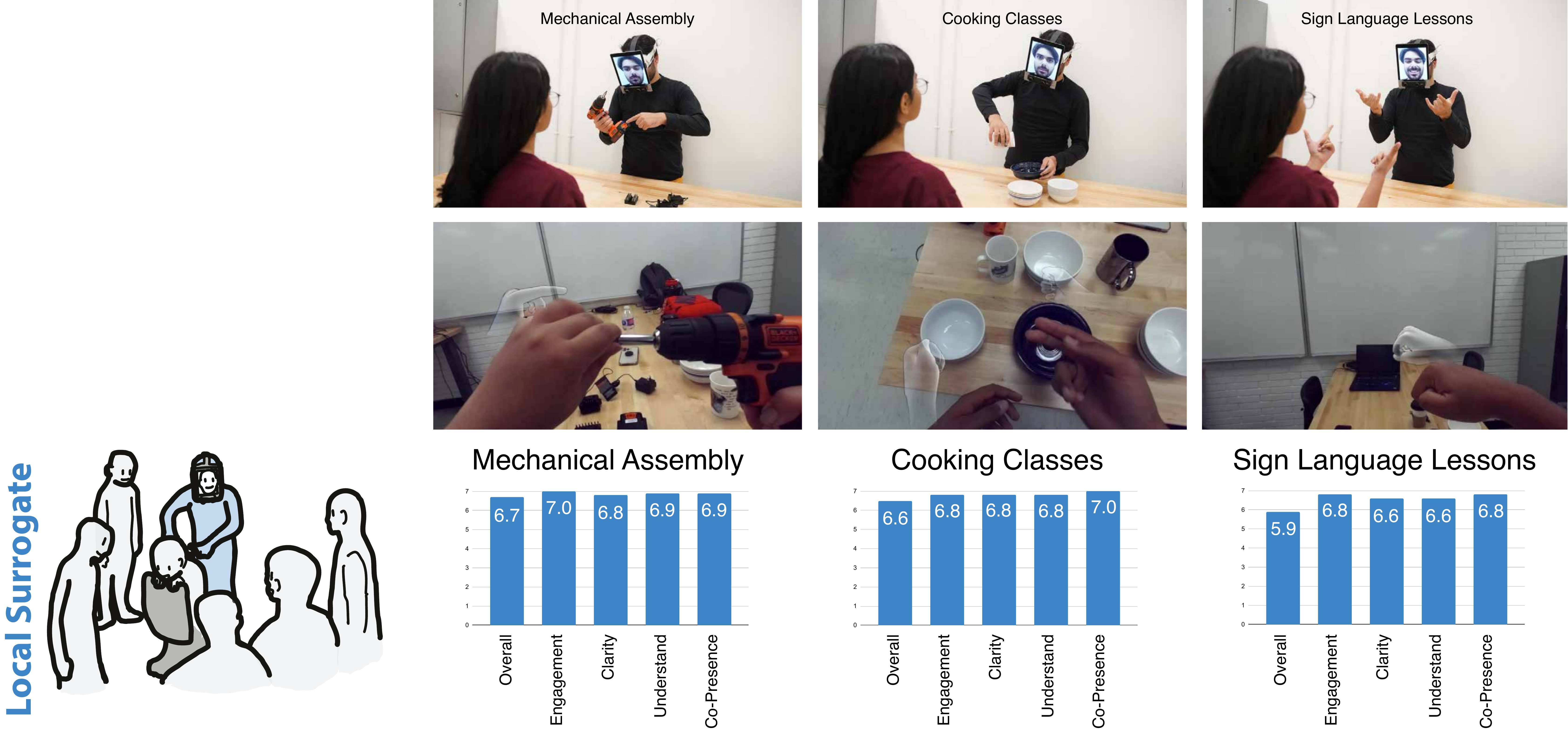}
\caption{Usability test for different (non-physiotherapy) applications and their results.}
\label{fig:non-massage-study}
\Description{This image displays a data visualization showing the results of the third user study which was conducted to assess ChameleonControl's suitability for non-physiotherapy scenarios. Namely: Mechanical Assembly, Cooking Classes, Sign Language Lessons.}
\end{figure*}

\ \\
\subsub{RQ4. Immersive vs Non-Immersive Guidance for Remote Instructors}
When comparing our mixed reality gestural guidance and the Zoom-based guidance, the instructors report our system is significantly \textbf{easier to control} (the surrogate) than both 1st POV and 3rd POV video-based guidance for remote instructors (Ours: 6.4, 1st POV: 4.1, 3rd POV: 3.6).
The instructors also report our system is significantly \textbf{easier to teach} than both 1st-person POV and 3rd-person POV video-based guidance (Ours: 6.1, 1st POV: 5.0, 3rd POV: 5.6).
For example, the instructors mention that 
\textit{``P1: Evaluating students is much easier with the AR system. With Zoom, you need to stop the demonstration and look at what students are doing. With the Ar system, I can help them in real-time.''}.
\textit{``P3: Student feedback is real-time, as is instructor feedback. This is much faster than Zoom.''}.

For other metrics, the instructors report positively about the overall experiences (Ours: 6.4, 1st POV: 4.6, 3rd POV: 5.8), the sense of co-presence (Ours: 5.2, 1st POV: 4.0, 3rd POV: 4.5), and clarity and informative (Ours: 4.6, 1st POV: 4.0, 3rd POV: 4.3).
For example, the instructors mention that 
\textit{``P1: Using Zoom, I need to practice on someone here. Using AR lets me demonstrate how the student needs to perform on their particular client. This makes me feel like I'm with the student.''}.
\textit{``P2: they were all similarly good, but I feel more engaged with the student using AR.''}.
\textit{``P3: I like getting instant student responses using AR. Wish I could feel their client tho''}.

\subsection{Usability Test for Different Applications}

\subsubsection{Research Questions}
For the third user study, our goal is to evaluate the \textbf{applicability to different domains and scenarios}. 
To achieve this, we designed our study with the following three application domains:

\begin{itemize}
\item \textbf{Sign Language Class}
\item \textbf{Cooking Lesson}
\item \textbf{Mechanical Assembly Lesson}
\end{itemize}
We chose these conditions because we are interested in evaluating the following research question: 
\begin{itemize}
\item \subsub{RQ5. Accuracy of Teleoperation for Different Applications} Does our mixed reality gestural guidance provide accurate and responsive teleoperation for different application scenarios?
\end{itemize}

\subsubsection{Method and Participants}
To evaluate our system, we recruited nine participants from the local community (3: female, 6: male; age: 18-23).
The study was conducted in a research lab. 
Participants were first provided an overview of our study, and then experience the four conditions one by one.
We held three different sessions with 3 participants each so that the order of running the conditions are randomized for each session.

The author acts as a remote instructor, and the participant is supposed to act as a local surrogate and synchronize their hands to the instructor's virtual hands. 
We capture with a video recording.
For \textbf{sign language}, we prepared eight American Sign Language (ASL) words \textit{Yes, No, You, Little, Sorry, How, Okay, Great} as well as a phrase \textit{What's up! How are you?} to each surrogate student.
For \textbf{cooking lessons}, we prepared for a mock cooking recipe using five small bowls and two utensils (knife and fork) to prepare a recipe by cutting, poking, shaking, and more. 
For \textbf{mechanical task}, we prepared for the disassembly tasks of the drill, which consists of five steps: \textit{setting drill safety switch, removing the battery pack, removing drill bit, removing drill bit holder, connecting the battery pack to a charger}. 
For each condition, the participant performed the task and after 5-10 minutes we switched to the next condition.
Once the participants experienced all conditions, they were then asked to fill out an online questionnaire.
In total, the study took approximately 40 minutes and each participant was compensated 10 CAD.

\subsubsection{Results}
Figure~\ref{fig:non-massage-study} summarizes the 7-point Likert-scale questions for each scenario.
Below, we discuss the following aspects in more detail.

\ \\
\subsub{RQ5. Accuracy of Teleoperation for Different Applications} 
When evaluating our mixed reality gestural guidance as local surrogates, participants rated our system generally well across each condition for each hands-on lesson, as shown in Figure~\ref{fig:non-massage-study}. More useful information came from participant comments detailed below.

Many participants mentioned that the mixed reality approach is easy to follow (P2, P7, P8, P9) and easy to understand (P5, P7). 
\textit{``P5: with the immersive ar lesson the instructor's hands were easy to trace and steps were easy to replicate. 1st person wasn't as immersive and 3rd person was just bad''}.
\textit{``P6: The immersive AR was very practical for me, the 1st person zoom gave me more of a personal point of view perspective, but I still preferred the hands-on experience with the instructor as in the AR Lesson''}.
\textit{``P8: The engagement was so good it was almost scary. It was like having a hallucination.''}. 
\textit{``P9: I could see how to position my hands After seeing my instructor's hands''}
\textit{``P2: It really helps as I don’t waste time looking to the screen and pausing the video or having the live stream going''}.
However, some participants complained of a lack of resolution and detail, an inherent limitation of the ZED Mini camera.
\textit{``P6: The vision wasn't very clear so it wasn't so it was a bit difficult to follow with cooking and drill disassembly. Sign language was very easy to follow''}.

\section{Discussion and Future Work}

\subsubsection*{\textbf{The Uncanny Valley Effect and See-Through Mixed Reality Headset for Local Surrogates}}
While surrogates themselves rarely complained about the face-mounted iPad, many participants in the student audience expressed that they found the face-mounted iPad to be ``weird'' or ``creepy''. 
One possible explanation for this is that the face-mounted iPad creates a mismatch between the remote instructor's face and the local surrogate's body. For example, the remote instructor may not always be displayed in the center of the screen, which can create an unrealistic feeling for students. In addition, in massage therapy classrooms, the surrogate is usually surrounded by students, making it hard for some students to see the instructor's face from the side view. This problem could be alleviated in lecture-style settings where the students can always see the face from the front.
Alternatively, another possible explanation is that the 2D iPad face may greatly invoke the \textit{uncanny valley effect}, which causes an increasingly negative emotional response when the subject becomes ``more human-like''. In this case, the poor replication of the surrogate's 3D face (more human-like) may cause a more negative feeling than the non-face condition (less human-like) due to the uncanny valley effect.
Participants claimed that they would likely prefer if the surrogate used a see-through mixed reality headset, such as the Microsoft Hololens, which would result in a more natural classroom experience and fewer distractions during the physical demonstrations.

\subsubsection*{\textbf{Ethics of Human Surrogate}} Something important to consider when using human surrogates in teleoperation systems is the distinction between \textit{voluntary} and \textit{involuntary} synchronization between surrogates and remote users. While \system{} can only function with a willing participant as the surrogate, other systems could be developed that blur this distinction. For example, using electro-muscular stimulation (EMS) technologies like ~\cite{lopes2015affordance++, tanaka2022electrical}, it is possible to influence surrogates to act or behave according to the will of the remote user. 
However, this raises ethical concerns about the potential for involuntary or forced synchronization, as well as the potential for abuse of such technology. 
Therefore, it is also important to consider the humanity of human surrogates in teleoperation systems. Replacing a surrogate's movements, behaviors, voice, body parts, or face with those of a remote user can risk dehumanizing the surrogate, particularly to the audience. 
On the other hand, it is possible to imagine a future where people are willing to be surrogates as work, similar to how people currently work as delivery drivers for services like Doordash. In such a scenario, there could be a crowdsourcing ecosystem around students, surrogates, and remote instructors that provides personalized, on-demand, hands-on education from anywhere in the world like \cite{suzuki2016atelier} but through a human surrogate system. 
However, this also raises ethical concerns about the potential exploitation of surrogates and the need for fair compensation and working conditions.
In any case, future developments in human-surrogate teleoperation should carefully consider the well-being, bodily autonomy and dignity of human surrogates.

\subsubsection*{\textbf{Ambiguity in Hand Gestures}}
While communication between surrogates and remote instructors was generally straightforward and clear, the current setup sometimes causes unclear ambiguity whether hand gestures were intended to communicate with the surrogate or to serve as an instruction to be imitated. 
For example, we observed that a remote instructor's thumbs up led to confusion for the surrogate who might unintentionally imitate the gesture. 
Despite this limitation of the system, we did not see any problems from the practical point of view, as the participants never complained about misunderstandings or miscommunication related to differentiating between gesturing for communication and synchronization.
Surrogates were able to discern when instructors wanted them to synchronize movements or gestures and when instructors were simply communicating with them. Verbal communication was also used to clarify any potential ambiguities in hand gestures. Overall, the use of hand gestures and verbal communication effectively reduced the potential for confusion or miscommunication.

\subsubsection*{\textbf{Outside-In Tracking for Whole Body Synchronization}}
In our system, we only focus on hand synchronization, but there is a clear opportunity for whole-body synchronization. Not only for physical therapy, but even for other applications such as sports training, dancing, martial arts, and more. Synchronizing more than just the hands would likely create a much more immersive and improved educational experience for the surrogate in addition to a significantly more accurate spatial and physical educational experience for the student audience. In particular, the surrogates, students, and instructors alike requested the ability to synchronize their legs, arms, torsos gaze, and head.
Future work should explore this whole-body synchronization through external depth cameras or motion tracking systems.

\subsubsection*{\textbf{More Accurate Hand Tracking and Head Movement}}
The remote instructor and local surrogates also complain about the accuracy of hand tracking. Implementing more comprehensive and accurate hand-tracking would likely result in further improvements to the educational experience and better spatial/physical demonstrations. Hand occlusion was also a problem since the instructor's hands would often inadvertently move behind an object due to the surrogate's movements since the virtual hand coordinates were bound relative to the surrogate's camera. Some instructors, particularly the physical therapists, expressed a strong desire to overlap their hands for particular techniques and lessons, which would usually occlude one hand completely resulting in a loss of hand-tracking. Increasing the trackable area for the instructor's hands would enable more flexible and natural teaching. Removing tracking devices from the instructor's forehead to fixed positions could provide an easier teaching experience. As stated previously, instructors universally requested an improved method to control the surrogate's gaze by incorporating the gaze visual cues.

\subsubsection*{\textbf{Haptic Feedback and Capturing Local Information for Remote Instructors}}
The instructors, in particular the physical therapists, mention that one major downside to \system{} is the loss of a major human sense: touch. To alleviate this, future research may want to explore the use of haptic feedback technologies. While previous work shows that the haptic feedback does not change the task completion time, it does show a significant increase in usability~\cite{wang2020haptic}.
By using haptic devices like haptic gloves~\cite{hinchet2018dextres} or ground-based encountered-type haptics~\cite{suzuki2021hapticbots} capturing the local user's touch sensation for the remote instructor, future research could further bridge the gap between remote and co-located hands-on education.
In addition to losing the sense of touch by using remote education technologies, other types of information are lost as well. Feedback from instructors indicated that instructors would like various real-time information sent from the surrogate. For example, the physical therapist instructors mentioned that measuring and relaying the pressure applied by the surrogate student could provide valuable real-time information for evaluating and correcting student techniques and behaviors.

\subsubsection*{\textbf{Long-term Learning Effects through In-the-Wild Studies}}
In this study, we did not evaluate any learning effects, such as pre-test and post-test analysis. The physio and massage therapy students were familiar with the instructions given by instructors, thus deployment with a fresh incoming group of students could reveal more about how systems like ours can impact students' educational experiences more accurately. Finally, to evaluate potential learning benefits, future work could explore long-term classroom use of real-human teleoperation systems like ours.

\section{Conclusion}
We present \system{}, a scalable, remote instruction system for hands-on classrooms using teleoperated real-human surrogates.
On top of the existing human-based telepresence~\cite{misawa2015chameleonmask}, this paper contributes to the first concept of \textit{real-human teleoperation} by combining human-surrogate telepresence and mixed reality gestural guidance.
By overlaying the remote instructor's virtual hands in the local user's MR view, the remote instructor can guide and control a local user, as if the remote instructor is teleoperating a real human. This enables the remote instructor to demonstrate hands-on, physical, and spatial lessons remotely and scalably to a classroom of students.
We deployed and evaluated our system in actual classrooms of massage therapy training, where 20 students evaluate and compare our system with video or mobile AR remote instruction from the student audience perspective. Additionally, 10 instructors and 14 students evaluate our system from the remote instructor and local surrogate perspectives and we conduct a third user study with 9 non-massage participants to evaluate our system for different (non-physiotherapy) applications, including sign language, cooking lessons, and mechanical assembly. The study results confirm that our approach can increase the engagement and sense of co-presence, showing the potential for the future of remote hands-on classrooms. We hope to inspire further research within real-human teleoperation.

\begin{acks}
This research was funded in part by the Natural Sciences and Engineering Research Council of Canada (NSERC RGPIN-2021-02857) and MaKami College. We also thank all of the experts and participants of our user study.
\end{acks}

\ifdouble
  \balance
\fi
\bibliographystyle{ACM-Reference-Format}
\bibliography{references}

\end{document}